# From Functions to Object-Orientation by Abstraction

*Bob Diertens*

section Theory of Computer Science, Faculty of Science, University of Amsterdam

*ABSTRACT*

In previous work we developed a framework of computational models for function and object execution. The models on an higher level of abstraction in this framework allow for concurrent execution of functions and objects. We show that the computational model for object execution complies with the fundamentals of object-orientation.

*Keywords:* programming, computational model, execution model, machine model, sequential execution, concurrency, object-orientation

## 1. Introduction

Mostly, an user's view on functions and objects stem from their use in programming languages. Although the use of functions and objects is explained with the programming languages, a computational model for them is not given. These models are implicit to the compilers of the programming languages and the machines the compiled programs are executed upon. In understanding the semantics of functions and objects and the relation between functions and objects knowledge of the computational models for them is essential.

Ever since object-orientation was introduced in programming languages there have been efforts in transforming programs built up from functions into object-oriented programs. All these efforts have in common that they lack an explicit descriptions and computational models for functions and objects.

In our view the execution of a function can be seen as the execution of an object. It is first instantiated from its definition, then executed, and then destroyed. Here, the instantiation and destruction are implicit, in contrast with the execution of an object where these are explicit. With a function, there is no control over its lifetime. An object has a state, and this state can be manipulated by invoking methods of the object (making requests to the object). This state is implicit with the object and exists only during the lifetime of the object. During several executions of a function a state can be preserved by explicitly storing it in memory.

The execution of a program built up from functions can be seen as the execution of an object. The functions are the methods of the object, and all the variables as the state of the object. A variable actually is itself an object. It is instantiated, manipulated by invoking certain operations on it, and destroyed.

We conclude that the program, its functions, and its variables are all objects. The methods of an object are in essence functions and are thus objects too. The state of an object can be seen as the variables of that object and are thus objects as well. So in essence there is not much difference between elements of function-oriented and object-oriented programs. The differences are determined by the computational models for them.

In previous work we described a framework of computational models that allow for the concurrent execution of functions ([1]. We extended this framework with communication between functions in [2]. In [3] we applied this framework to object execution, resulting in a framework of computational models that allow for the concurrent execution of objects. The sequential execution of functions and objects is just a possible implementation of an abstract computational model that allows for concurrent execution. This work clearly supports our conclusion that everything can be considered an object.

Although everything can be considered an object, this does not mean we can considered it complying with the object-orientation paradigm. For this it has to fulfil the fundamentals of object-orientation that we described in previous work [4]. There we defined a model for object-orientation with as much abstraction as possible so that it can be applied in several phases of software development. In this article we show that, through the abstractions we have made to establish the framework of computational models, we comply with the fundamentals of object-orientation.

In section 2 we describe our previous work on computational models for function and object execution. We summarize our view on object-orientation in section 3. And in section 4 we show that the computational models comply with the fundamentals of object-orientation.

## 2. Computational models

In this section we give an overview of previous work on computational models for function execution [1], communicating concurrent functions [2], and computational models for object execution [3].

*2.1 Models for function execution*

In [1] we developed a framework of computational models at different levels of abstraction for the concurrent execution of functions. The development of the framework started with the traditional sequential execution model for functions from which a sequential computational model was obtained by abstracting from the details of function call implementation. Further abstraction from the way a function is scheduled for execution led to an abstract computational model that allows for the concurrent execution of functions (Figure 1).

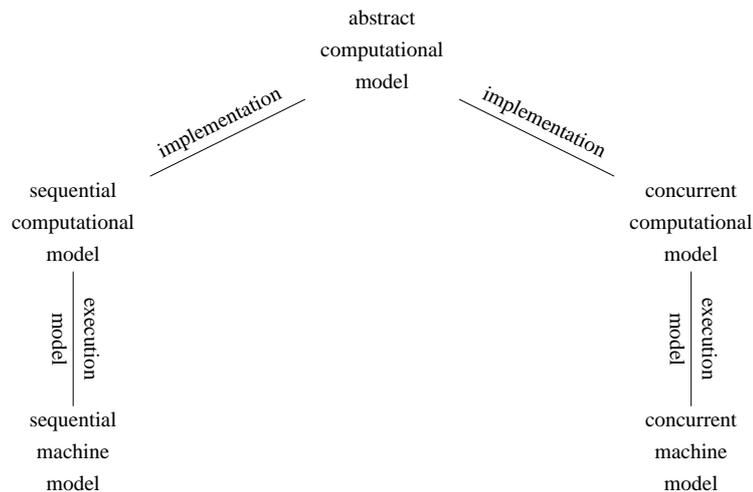

**Figure 1.** Framework of computational models

It showed that with abstraction and relaxing constraints a model for execution of functions can be obtained in which function scheduling plays a key role. This model has as a possible implementation inline scheduling, the original stack-based function execution model the development of the framework started with. But more important, this model allows for the concurrent executions of functions, and therefore it can be used as model for the implementation of concurrent software. The framework of computational models at different levels of abstraction can be used for further development of concurrent computational models that deal with the problems inherent with concurrency.

*2.2 Communication between functions*

Concurrent execution of code allows for communication between the parts of code that are executed concurrently. The communication can be done either through models using shared memory or by models using message passing. With message passing models communication is done by exchanging messages and shared memory communication is done by reading from and writing to memory that can be accessed by at least all parties in the communication. In [2] we have extended the abstract computational model from our framework of computational models for the concurrent execution of functions with communication between the concurrent functions. For this, we classified the communications between concurrent functions in direct and indirect and in unidirectional and bidirectional. For the different kinds of communication we identified problems that can occur with these communications and we presented solutions for these problems. The solutions are based on encapsulating the area in which the problem occurs together with a mechanism that solves the problem. The encapsulation makes it possible to abstract from the innerworkings. The resulting mechanism can then easily be replaced with something else with the same behaviour.

We defined the basic locking mechanism that we use in our solutions (Figure 2). This mechanism itself is based on the locking mechanism commonly used in software, but through encapsulation and abstraction this is completely concealed.

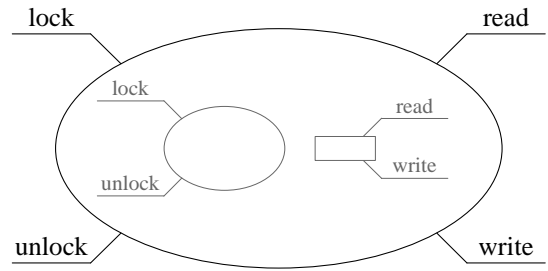

**Figure 2.** Indirect communication with encapsulated location and locking mechanism

For unidirectional indirect communication we defined the basic communication mechanism (Figure 3) that encapsulates the basic locking mechanism.

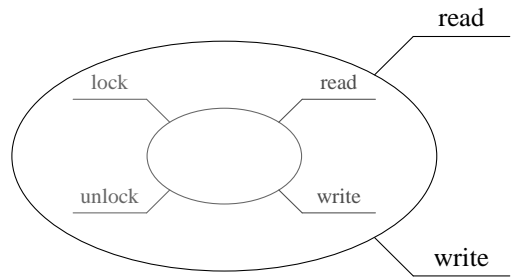

**Figure 3.** Indirect communication with encapsulated locking mechanism

This basic communication mechanism is constrained into the status based communication that implements unidirectional indirect communication (Figure 4). The constraints are based on the states of the mechanism that indicate whether a read or a write is possible.

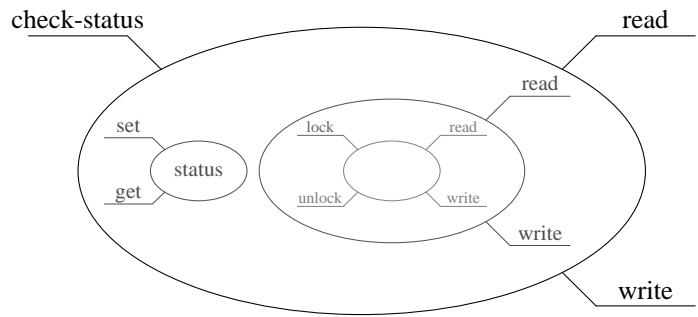

**Figure 4.** Unidirectional indirect communication with encapsulated status flag

Bidirectional indirect communication consisting of two unidirectional indirect communications that share the location for storing the messages in both directions can be implemented using a similar status based communication. But then the solutions is a bit more complex because there are more possible states. Not only is it necessary to indicate whether a read or a write is possible, but also by which party involved in the communication.

Furthermore, we discussed some special cases of communication. These cases can be implemented using one of the defined mechanisms, or similar implementations with the same behaviour, depending on the constraints that have to be enforced on the communications. A particular intriguing case is undirected communication what on a low level of abstraction is known as shared memory communication. This can be implemented with the basic communication mechanism or the basic locking mechanism when control over the locking mechanism itself is required.

When we abstract from the inner working of the mechanisms, these are just concurrent functions among all the other functions. The form of communication with these mechanisms is unidirectional direct communication. So we can conclude that, through our mechanisms, unidirectional indirect communication is built up from unidirectional direct communications.

*2.3 Computational models for objects*

The traditional sequential execution model for objects is based on the traditional sequential execution model for functions. In this model it seems that the objects are not executed, but the methods of the objects, and in a similar manner as functions are executed. We could therefor use the same framework of computational models for the methods of objects as the one developed for functions (Figure 1). This framework then allows for the concurrent executions of methods.

We can extend the framework so that it allows for the concurrent execution of objects. For this we have to see that the scheduler for the methods has to make a context switch to the object a method belongs to. In this framework the context switch is done inline with the scheduling of a method, similar to the inline scheduling of functions in the sequential computational model for functions. Apparently, the context switch is the scheduling of the object. Separating the scheduling of objects from the scheduling of methods gives us a scheduler for objects where each object takes care of the scheduling of its methods.

With this, we abstracted from how objects are scheduled. The result is a computational model that allows for the concurrent execution of objects. In this model, each object decides how to schedule its methods. Of course, the scheduling of methods can be inline and so limiting concurrency to objects only. From this model we can implement the abstract computational model that allows for the concurrent execution of functions or methods, but we can also implement a model that allows for the concurrent execution of objects in which each object decides how to schedule its methods.

What has been a call of a method for a particular object in the sequential execution model has now become a request to that object. How the object handles the request is not important, and so we can abstract from this.

## 3. On object-orientation

In [4] we described our view on object-orientation (OO) in software design of which we repeat here the main part of this paper.

OO is a modelling paradigm for describing objects and their relationships. Objects and relations are supposed to stand close to real world concepts. The real world is the world we are implementing, that is a level of abstraction in the design or a requirements specification. The real world is a future world in which the system under development takes part.

The real world is also an abstract world. It is of no concern how something works, only what it does. This abstraction is key in OO. However, OO is often easily replaced with object-oriented programming (OOP). But an implementation in an OOP language is no more than an example of this modelling at the lowest level of abstraction of the design.

Because of this replacement, OO is explained by describing what an particular OOP language has to offer. To define a model of OO that can be applied in several phases of software development we have to define this with as much abstraction as possible. We give a description of the fundamentals of OO, techniques to support the fundamentals, and features based on the techniques. Note that only the fundamentals are necessary for object-oriented modelling, some support can be nice, and features are mostly only used on the lowest levels of abstraction.

*3.1 Fundamentals*

OO can be seen as a kind of technique of organizing a system in terms of objects and their relations. It is supposed to stand closer to the *real* world as opposed to techniques predating OO. Its characterist is the distinction between the observable behaviour of objects and the implementation of the behaviours.

**Objects**
An object has the following characteristics.

> **state**
>> for recording the history of an object upon which future behaviour can be based.
>
> **behaviour**
>> the observable effects based on its state and the relations with other objects.
>
> **identity**
>> as known by other objects, either by name or by reference.

**Relations**
Relations between objects are expressed by interactions in the form of message passing.

**Abstraction**
Manipulation of an object can only be done through its relations with other objects. Thereby hiding the implementation of its behaviour and the recording of its state. It is only important what an object does, not how an object does it.

*3.2 Support*

An object-oriented language for modelling systems on a particular level of abstraction has to support the fundumentals of OO and possibly even enforce these fundamentals. Support can be provided in the following forms.

**Types**
An object type is a container in which the state and the behaviour(s) for an object are defined.

**Message Passing**
The way messages are passed between objects can be supported in more than one form.

**Encapsulation**
Encapsulation prevents objects from relating to each other in other ways then the provided forms of message passing.

**Information Hiding**
Hiding of information about an object can be done by deliberately making this information inaccessible.

*3.3 Structures*

Based on the techniques supporting OO structures can be formed. Such structures behave as objects themselves, characterizing the concepts of OO.

**Type Composition**
The basic idea of composition is to build complex object types out of simpler ones. Besides that objects can be built up from ways to define state and behaviour as provided by the modelling language, objects can also be built up from other object types. The latter can be done in the following forms.

> **reference**
>> An object type can reference an object of a particular object type.
>
> **inclusion**
>> An object type can include another object type.

To obey the OO fundamentals of keeping behaviour and implementation of an object distinct, a modelling language has to hide the composition of an object type. This can be achieved by making the elements of the object type acquired through composition available either only from within the object type, or from outside the object type but as it were elements of the object type itself.

Objects composed in this way are vertical related with the objects they are composed of.

**Object Composition**
Several inter-related objects form a cluster that when abstracted from the inter-relations acts as a single object. Objects that take part in this composition are horizontal related with eachother.

**Abstract Object Types**
An abstract object type is an object type described in terms of objects representing elements of the abstract object type for which the type(s) have to be filled in on a lower level of abstraction. An abstract object type can also be turned into a generic object type on a lower level of abstraction, with parameters for the object types representing the elements.

## 4. Compliance with fundamentals of object-orientation

In this section we relate the computational models from section 2 with the fundamentals of object-orientation described in section 3.1. We show how we can comply with these fundamentals, thereby linking our framework of computational models with the object-orientation paradigm.

**Objects**
The framework of computational models for function execution and object execution showed that, by abstracting from the ways of scheduling, functions can be considered objects. However, in order to make a shift to the paradigm of object-orientation these objects must have some characteristics as well. First they have to have to an identity by which they are known to other objects. Instances of functions are known by their identity obtained on creation of the instance. This identity can be passed on to other instances of functions. Secondly, they have to have a state, although this state may be empty. The instance of a function can keep a state with the use of local variables. It is also possible that an instance of another function is used to keep a state. And thirdly, they must have an observable behaviour. Their behaviour is observable by the effects they have on their environment based on their state and the relations with other objects. With these three characteristic an instance of a function can be considered an object in the modelling paradigm of object-orientation.

**Relations**
In section 2.2 we described communication mechanisms for concurrent functions that solve the problems inherent with concurrency when communicating through shared memory. These mechanisms are functions themselves and thus objects. By restricting communication only through these mechanisms, objects can only interact with each other by passing messages. The passing of messages between objects describes the relations between the objects.

**Abstraction**
The behaviour of a group of objects that is related to objects outside of the group only through one of its members can be observed as the behaviour of that one object. What is going on inside this group is not observable to objects outside of this group. Because of this, the internal behaviour of this group can be abstracted from. So, by restricting communications between instances of functions through the mechanisms that are instances of functions themselves, it is possible to hide the implementation of its behaviour and the recording of its state.

## 5. Conclusions

We showed how to get from the sequential execution of functions to object-orientation by abstraction. First, we abstracted from the way function calls are implemented. This led to an abstract computational model that allows for the concurrent executions of functions. We extended this model with communication between the concurrent functions. For this, we abstracted from the communication through shared memory. Communication between functions became message passing with mechanisms that are just concurrent functions among all the other functions. We applied our model to object execution by abstracting from the way the methods of the objects are scheduled. For this, we separated the scheduling of the objects from the scheduling of the methods of the objects. The result is a computational model that allows for the concurrent execution of objects in which each object decides how to schedule its methods. We showed that this computation model for object execution that allows for concurrency complies with the fundamentals of object-orientation.